# RUN UP OF SURFACE AND INTERNAL WAVES

*H. Branger*[1] *, O. Kimmoun*[2] *, N. Gavrilov*[3] *, V. Liapidevskii*[3] *, E. Pavlova*[4]
[1] *IRPHE, CNRS, Marseille, France ; e-mail : branger@irphe.univ-mrs.fr*
[2] *ECM, Marseille, France ; e-mail : olivier.kimmoun@ec-marseille.fr*
[3] *LIH, SB RAS, Novosibirsk, Russia ; e-mail : liapid@hydro.nsc.ru*
[4] *POI, FEB RAS, Vladivostok, Russia : epavlova@poi.dvo.ru*

The evolution of breaking waves propagating towards the shore and more specifically the run-up phase over the swash-zone for surface as well as for internal waves is considered. The study is based on the laboratory run up experiments for surface waves performed in ECM (The Ecole Centrale Marseille), on the laboratory stratified flow experiments performed in LIH (Lavrentyev Institute of Hydrodynamics) and on the field data describing the internal wave run up from the joint LIH – POI (V.I. Il'ichev Pacific Oceanological Institute) expedition in summer 2006. The presentation is focused on the breaking and energy transfer mechanisms common to surface and internal waves in the swash-zone. The mathematical model taking into account turbulent mixing and dispersion effects is discussed.

**Run up of large amplitude breaking waves**

Experiments have been conducted in the glass-windowed ECM wave-tank (17m x 0.65m x 1.5m) in Marseille. A 1/15 sloping beach is mounted on the bottom. The wave maker is able to generate regular waves, freak waves (by space-time focalization technique), or valuable solitary waves. Wave profiles have been measured by a set of wave gauges, velocity measurements have been performed by Particle Image Velocimetry (PIV) methods. Synchronized measurements are done at different locations and the space/time velocity field are reconstructed over all investigated zone. The extension of the wave run up, and the celerity of the propagating bore on shore are estimated using video techniques with a vertical laser light sheet. The details of the experimental procedure can be found in Kimmoun et al. (2004).

Much attention is given to the correct synchronization between experiments to have a global estimation of the mean and fluctuating parts of the wave induces velocity flow. Different types of waves have been investigated: regular waves, groups of waves, extreme waves and solitary waves. Different stages of run up process from a spilling breaker to the developed turbulent bore are shown in Figure 1.

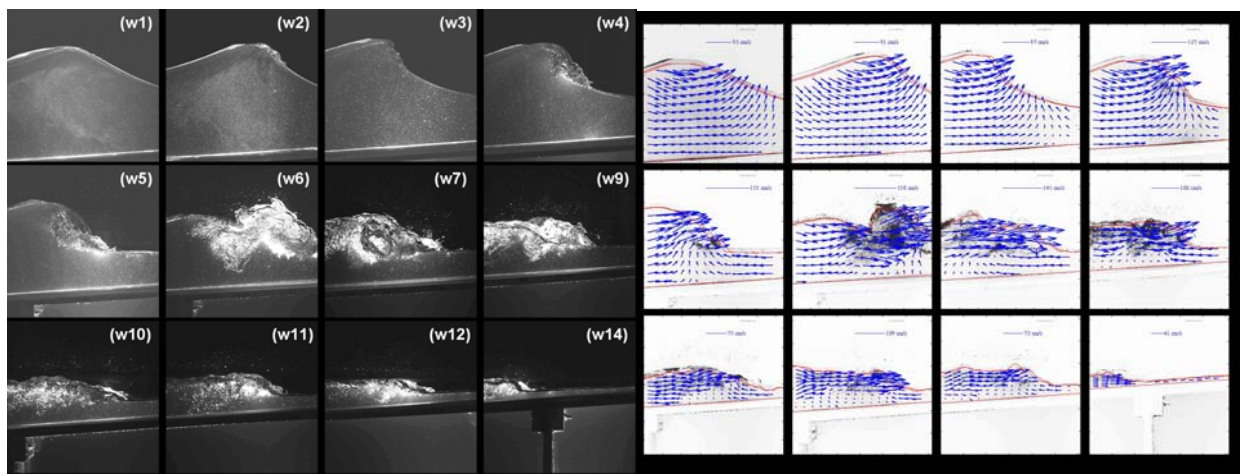

Figure 1. Run up of surface waves

**Large-amplitude internal waves in shelf zone**

The topic of internal waves is studied in the context of wave breaking, vortex formation and mixing in coastal waters. Nonlinear internal waves generated by tides, as well as by interaction of flows with topography, play an important role in the energy transfer from the large-scale motion to small-scale mixing. The analogy with the breaking mechanisms of surface waves is very useful to simulate the dissipation processes on shelf. Vertical and horizontal structure of near-shore stratified flows is important since some abnormal wave shapes can occur close to the breaking limits. Scotti and Pineda (2004) have recently observed bore-like structures with trapped cores in the near shore area of Atlantics. The transition from wave-like motion to the separate moving soliton-like structures containing trapped dense core is the common feature of the run up process of internal waves. It can be observed in any shelf zone with high internal wave activity as well as in laboratory experiments. In figure 2 the run up of lenses of cold water due to interaction of the internal tide front with the bottom is shown.

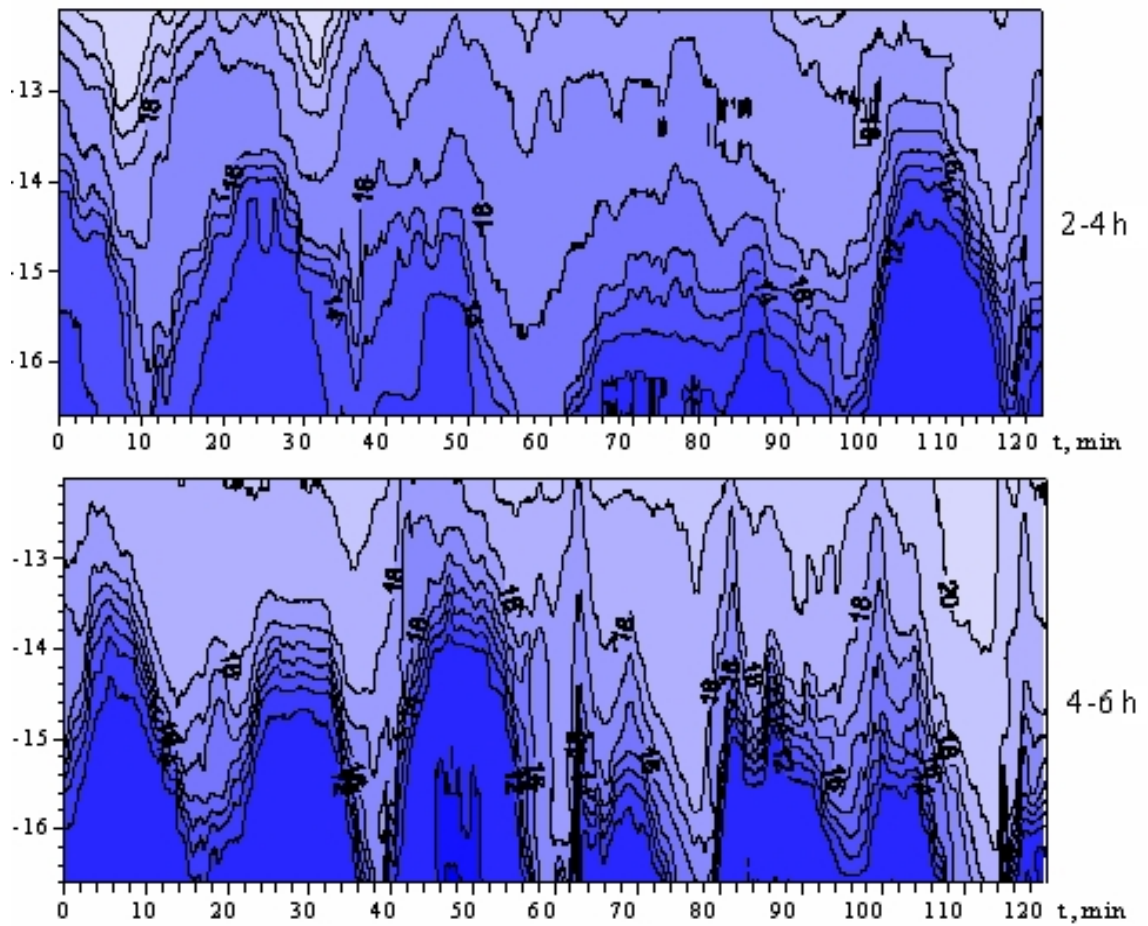

Figure 2. Run up of internal waves in the shelf. The total depth is 16.5 m, the temperature difference between the cold core and surrounding fluid is about 10 degrees.

The undisturbed depth of the thermocline is 25 meters. The run up phase of internal tide moves cold water to the shore for a long distance. Since the lenses of cold water are separate, the mean velocity of fluid in such cold cores is equal to the phase velocity of waves. The dissipation due to mixing is very effective in this region too. It means that run up of internal waves influences many dynamical processes in coastal waters (sediment transfer on the shelf, vertical mass exchange, etc).

The laboratory experiments on internal wave run up performed in the glass-windowed LIH wave-tank (3m x 0.15m x 0.6m) in Novosibirsk reveal the dependence of mixing efficiency on the thickness of dense layer before the wave front. In Figure 3 the stages of run up process of an internal wave is shown. One can see the sudden transition from a laminar flow to the high turbulent flow in the core of wave, when the depth of the dense layer before the wave vanishes.

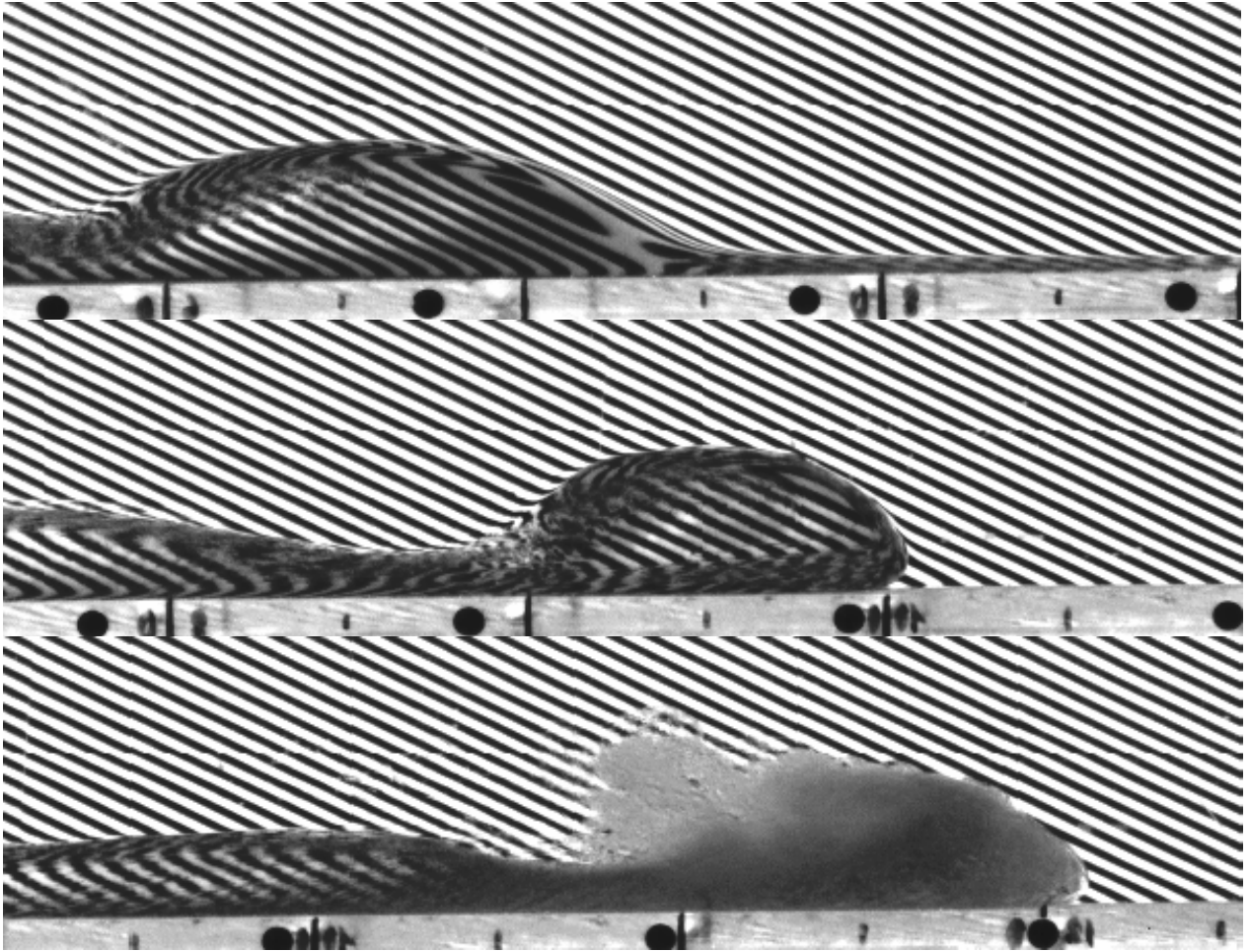

Figure 3. Run up of an internal wave (LIH experiment): the soliton-like wave (upper), the bolus formation at the boundary of the swash-zone (middle), the turbulent bore on the "dry bottom".

The mathematical model, which is an extension of the Green-Naghdi model and includes the turbulent layer, has been developed to find the criterion of wave breaking in the steady-state flow over a topography (Liapidevskii and Xu, 2006). This model and its hyperbolic approximation is adopted to unsteady run up problems for surface and internal waves.

The work was supported by INTAS under grant 06-9236, by RFBR under grant 05-05-64460, by RAS under grant 4.13.1 and by SB RAS under grant 06-113.

**References**

Kimmoun O., Branger H., Zucchini B. (2004) "Laboratory PIV Measurements of Wave Breaking on Beach" // ISOPE , Toulon, France, paper 2004-VR2-02, 6pp.

Liapidevskii V. & Xu J. (2006) "Breaking of waves of limiting amplitude over an obstacle"
 // Journ. Appl. Mech. Techn. Phys., vol. 47 (3), 307-313.

Scotti A., Pineda J. "Observation of very large and steep internal waves of elevation near the Massachusetts coast" // Geophysical Res. Lett. V31, L22307